\documentclass[twoside]{article}

\usepackage{graphicx}

\usepackage{cmpj2e}

\title[Classical and quantum antiferromagnets]%
{Classical and quantum anisotropic Heisenberg antiferromagnets}%
\author[W. Selke et al.]{W. Selke\refaddr{label1},
  G. Bannasch\refaddr{label2},
  M. Holtschneider\refaddr{label1}, I.P. McCulloch\refaddr{label3},
  D. Peters\refaddr{label1}, and S. Wessel\refaddr{label4}}
\addresses{
\addr{label1} RWTH Aachen University and JARA-SIM, Department of Theoretical
Physics, 52056 Aachen, Germany
\addr{label2} MPI f\"ur Physik komplexer Systeme, 01187 Dresden, Germany
\addr{label3} University of Queensland, Brisbane, QLD 4072, Australia
\addr{label4} University of Stuttgart, Institute for Theoretical Physics III, 70550 Stuttgart, Germany}
\begin{document}

\maketitle

\begin{abstract}
We study classical and quantum Heisenberg antiferromagnets with exchange
anisotropy of XXZ--type and crystal field single--ion terms of quadratic and
cubic form in a field. The magnets display a variety of phases, including the
spin--flop (or, in the quantum case, spin--liquid) and
biconical (corresponding, in the quantum lattice gas description,
to supersolid) phases. Applying ground--state
considerations, Monte Carlo and density matrix renormalization
group methods, the
impact of quantum effects and lattice dimension is analysed. Interesting
critical and multicritical behaviour may occur at quantum and
thermal phase transitions.
\pacs 05.10.Ln, 75.10.Jm, 75.40.Mg, 75.40.Cx
\end{abstract}

\section{Introduction}

Uniaxially anisotropic Heisenberg antiferromagnets in a magnetic
field along the easy axis have been studied quite extensively
in the past, both experimentally and
theoretically \cite{aha,shap}. Typically, they display in the
ground state and at low temperatures, the antiferromagnetic (AF)
and, when increasing the field, the spin--flop (SF) phase, as
depicted in Fig. 1. 

\begin{figure}
\centerline{\includegraphics[width=0.65\textwidth]{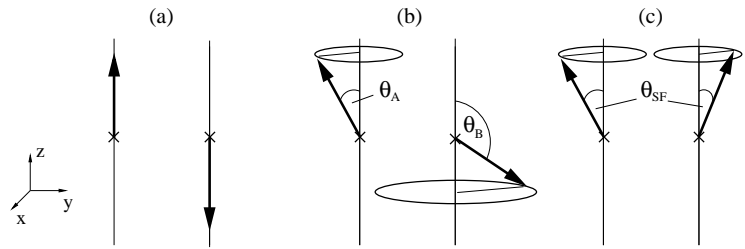}}
\caption{Spin orientations on neighboring sites showing perfect
     antiferromagnetic (a), biconical (b), and spin--flop (c)
     structures.}
\label{eps1}
\end{figure}

There may also occur intermediate or mixed spin structures
of 'biconical' \cite{mef} (BC) type. They are characterized by
distinct turning angles, $\Theta$$_A$ and $\Theta$$_B$, with respect
to the easy axis for neighboring spins situated on the two
sublattices, see also Fig. 1. Quite a few experiments related to BC
structures have been performed over the years \cite{old,bog}.

Recent studies deal with the influence of
defects \cite{sel,leidl}, of the lattice dimension,
of details of the anisotropic interactions, and of quantum effects
on the phase diagrams. In particular, topological and (multi)critical
properties of the phase diagrams have been elucidated.

The aim of the present contribution is to draw attention to recent 
analyses on the prototypical XXZ Heisenberg antiferromagnet in
a field and variants. The XXZ model is assumed to have a
uniaxial exchange anisotropy, with the magnetic field being
along the easy axis. By then adding crystal--field
single--ion terms of quadratic or cubic form, favoring
non--uniaxial spin orderings, biconical structures
may be stabilized. Much actual
interest in these BC structures stems from the fact that they
correspond to supersolid structures, when transcribing the
Heisenberg antiferromagnet to a quantum
lattice gas \cite{mat,tsu,liu,novi}. We shall consider
both classical models, with spinvectors of length unity, and
quantum models, with spin S=1/2 and S=1. In addition, to discuss
the role of lattice dimension, results for chains, square
lattices, and cubic lattices will be presented. 

\section{XXZ antiferromagnets}

As a starting point of theoretical studies on uniaxially anisotropic
Heisenberg antiferromagnets, one often considers the
prototypical XXZ model, with the Hamiltonian

\begin{equation}
  {\cal H}_{\mathrm{XXZ}} = J \sum\limits_{i,j}
  \left[ \, \Delta (S_i^x S_j^x + S_i^y S_j^y) + S_i^z S_j^z \, \right]
  \; - \; H \sum\limits_{i} S_i^z
\end{equation}

\noindent
where $J > 0$ is the antiferromagnetic exchange coupling between spins
being located on neighboring lattice sites $i$ and $j$. $\Delta$
is the exchange anisotropy, $1 > \Delta > 0$ in the case
of uniaxiality, and $H$ is the applied
magnetic field along the easy axis, the $z$--axis. 

In the following, we shall discuss the phase diagrams of the XXZ 
antiferromagnets for square and cubic lattices, showing similar
as well as quite distinct features. The phase diagrams have 
been obtained using mainly Monte Carlo (MC) simulations applying
the Metropolis method for classical models and applying stochastic series
expansion techniques for quantum systems, augmented by
finite--size analyses. 

In our MC simulations we studied two--dimensional
systems with up to $240^2$ classical and up to $150^2$
quantum, S=1/2, spins. In three dimensions, we simulated 
systems with up to $32^3$ classical spins. To estimate
error bars, we averaged over several rather long runs. In all cases, full
periodic boundary conditions have been employed.  

\subsection{Square lattice}

\begin{figure}[h]
  \includegraphics[width=.45\textwidth]{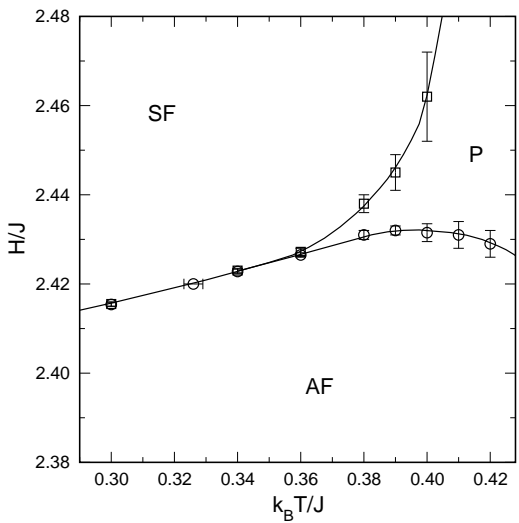}
  \caption{Phase diagram of the classical XXZ model on a square 
    lattice near the maximum
    of the boundary line of the AF phase, $\Delta= 4/5$, as obtained
    using Monte Carlo simulations.}
  \label{eps2}
\end{figure}

The {\it classical} XXZ model on a square lattice is known to display
in the (temperature T, field H)--plane
ordered AF and SF phases \cite{land,holt1,zhou,holt2}, see Fig. 2. The
transition to the paramagnetic phase
belongs to the Ising universality class along the boundary
line of the AF phase, and to the XY (or
Kosterlitz--Thouless) universality class along the boundary
line of the SF phase.

\begin{figure}[h]
  \includegraphics[width=.5\textwidth]{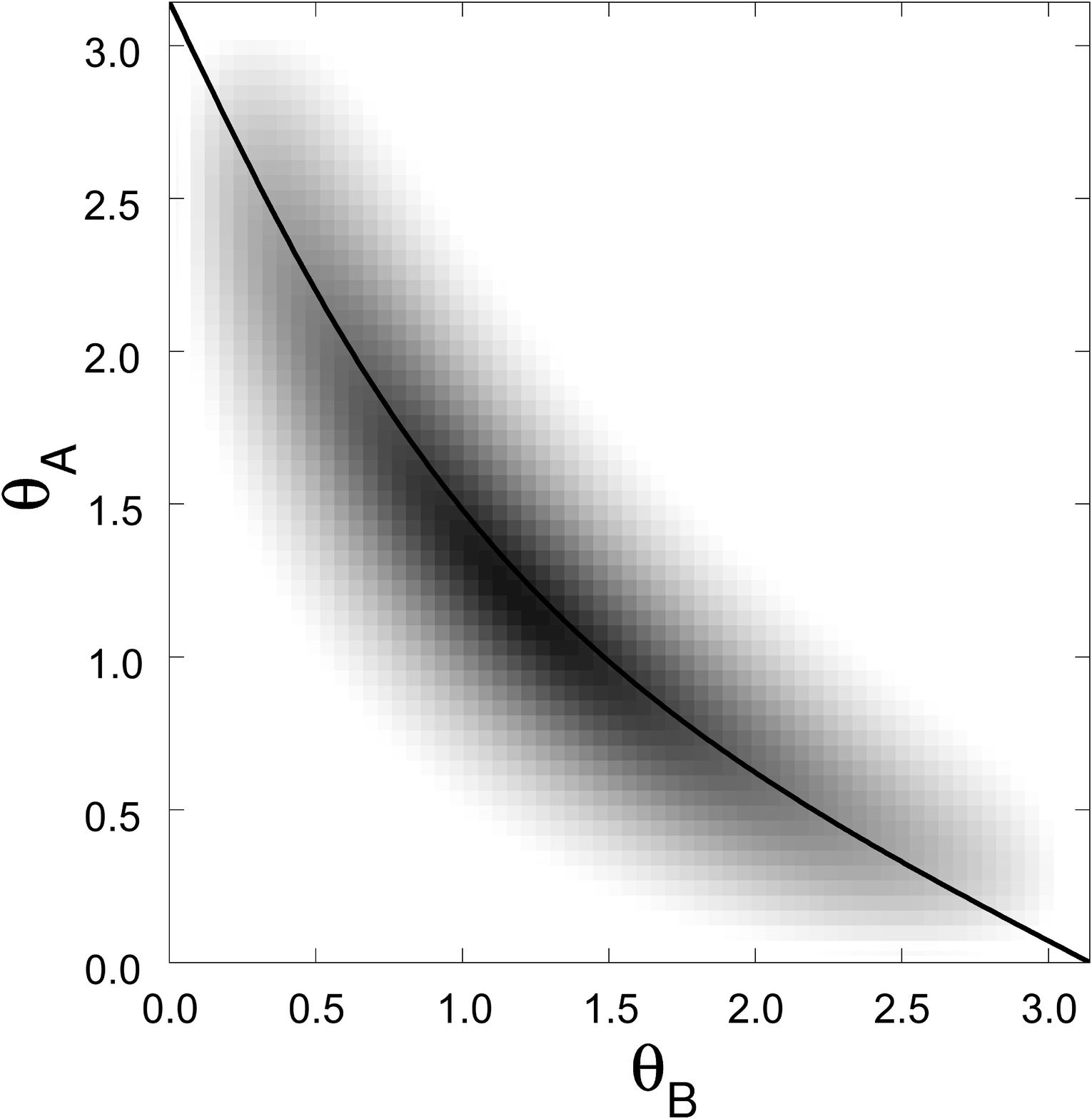}
  \caption{Probability ~\mbox{$p_2(\Theta_A,\Theta_B)$}
    showing the correlations between the tilt angles~$\Theta_A$
    and~$\Theta_B$ on neighboring sites for the
    XXZ antiferromagnet on a square lattice, with~$80 \times 80$
    spins, at~$H/J=2.41$, $k_BT/J=0.255$,
    and~$\Delta=\frac{4}{5}$. $p_2$~is proportional to
    the gray scale. The
    superimposed black line depicts the relation between the two
    angles $\Theta_A$ and $\Theta_B$ in the ground state, eq. (2).}
  \label{eps3}
\end{figure}

Extensive Monte Carlo simulations
augmented by finite--size analyses \cite{holt1,zhou,holt2}
suggest that there is no direct transition between the AF and SF
phases, albeit the two ordered phases approach each other
very closely at low temperatures, see Fig. 2. The intervening, extremely
narrow paramagnetic (P) phase and the two ordered phases arise
from the highly degenerate ground state occurring at the field
$H_c = 4J \sqrt{1- \Delta^2}$.  At that point, not only
AF and SF configurations have the same energy, but also
biconical structures. The BC structures, Fig. 1(b), may be described
by the tilt angles $\Theta_A$
and $\Theta_B$, being uniquely interrelated by \cite{holt2}

\begin{equation}
 \Theta_B = \arccos \left( \frac{ \sqrt{1-\Delta^2} \; - \; \cos\Theta_A }{ 1 \; - \;
 \sqrt{1-\Delta^2} \cos\Theta_A } \right)
\end{equation}

\noindent
where the tilt angle $\Theta_A$ ranges from 0 to $\pi$. Indeed, the
degenerate BC configurations have been argued to lead
to the intervening disordered
phase between the AF and SF phases \cite{holt2}. Their 
presence at low temperatures may be demonstrated conveniently by monitoring
the probability $p_2(\Theta_A,\Theta_B)$ to find the two tilt
angles at neighboring sites, as illustrated in Fig. 3. Similar
observations hold for the anisotropic XY antiferromagnet on a square
lattice \cite{holt3}.

\begin{figure}[h]
  \includegraphics[width=.5\textwidth]{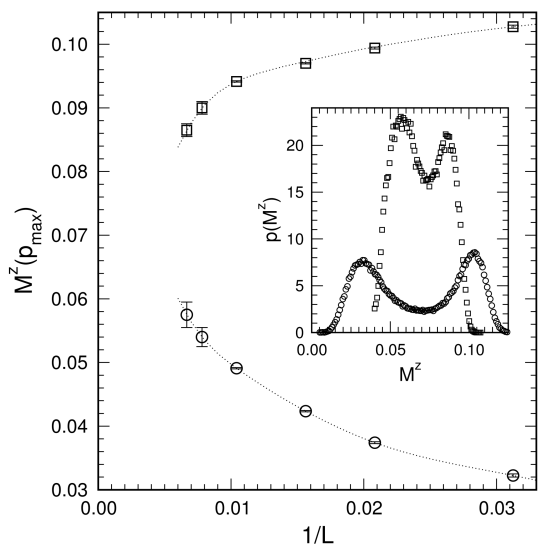}
 \caption{Positions of the maxima of the magnetization histograms as a
    function of the inverse system size, $1/L$, for the S=1/2 XXZ
    antiferromagnet on a square lattice, with $\Delta= 2/3$. The inset
    exemplifies two histograms
    for systems of linear size $L=32$~(circles) and $L=150$~(squares)
    at~$k_BT/J=0.13$ and the coexistence fields~$H/J=1.23075$
    and~$H/J=1.232245$ \cite{holt2}.}
\label{eps4}
\end{figure}

The degenerate
point at ($T=0,H_c$) may be coined a 'hidden
tetracritical point' \cite{holt2,zhou2}.

Due to the degeneracy, various physical quantities take
at $T=0$ and $H_c$ nontrivial
values, depending continuously on $\Delta$. For instance, the
magnetization per site is approximately 0.22 for $\Delta= 0.8$ \cite{holt3}.

In contrast, the {\it quantum} XXZ antiferromagnet with S=1/2 
on a square lattice seems to exhibit a direct
transition of first order between the AF and SF phases
\cite{schmid}. The triple point, at which the AF, SF and P
phases join, has been argued to be a critical endpoint, with the
boundary line between the AF and P phases being of first
order close to that triple point \cite{schmid}. The evidence 
had been provided mainly by quantum Monte Carlo simulations. We did
quantum Monte Carlo simulations as well, for the same anisotropy, namely
$\Delta =2/3$, using stochastic series
expansions with directed loop updates \cite{sand}, enlargening
the system sizes and improving
the statistics of the simulations in comparison to
the previous Monte Carlo study \cite{holt2,schmid}. Our results
suggest that the location of that special point
may be shifted towards lower temperatures, compared
to the previous estimate \cite{holt2}. This is illustrated in
Fig. 4, where very accurate MC data for various system 
sizes are shown at a temperature closely above that of the proposed
critical endpoint, where the previous study suggested a transition of
first order between the AF and P phases. Monitoring the size
dependence
of the peak positions in the magnetization
histograms, Fig. 4, the two peaks, may well coincide 
in the thermodynamic limit, signaling a continuous
transition \cite{holt2}. Because we do not find any
evidence for BC structures in our simulations of the quantum model,
there seems to be no mechanism for destroying long--range order 
at low temperatures in the quantum case. Thence the direct transition
between the AF and SF phases seems to be allowed, in
accordance with recent quantum MC findings \cite{wes}. In
general, quantum fluctuations may substantially reduce BC structures, compared
to the classical case, as will be discussed below.

\subsection{Cubic lattice}

The classical XXZ antiferromagnet on the cubic lattice displays also ordered
AF and SF phases, see Fig. 5. In agreement with previous
simulations \cite{bind}, we find the transition between the AF and SF phases
to be of first order \cite{ban}. There is again a highly degenerate
ground state, now at $H_c = 6J \sqrt{1- \Delta^2}$, with
BC configurations as described by eq. (2). Indeed, by
monitoring the probability $p_2(\Theta_A,\Theta_B)$, BC
configurations are observed to contribute to the thermal flucuations at
low temperatures. But, close to the
transition between the AF and SF phases, $p_2$ shows
pronounced local maxima corresponding to these two phases, indicating
phase coexistence at a transition of first order. In
addition, analysis of other standard thermodynamic quantities 
give evidence for a transition of first order \cite{ban}.

\begin{figure}[h]
  \includegraphics[width=.6\textwidth]{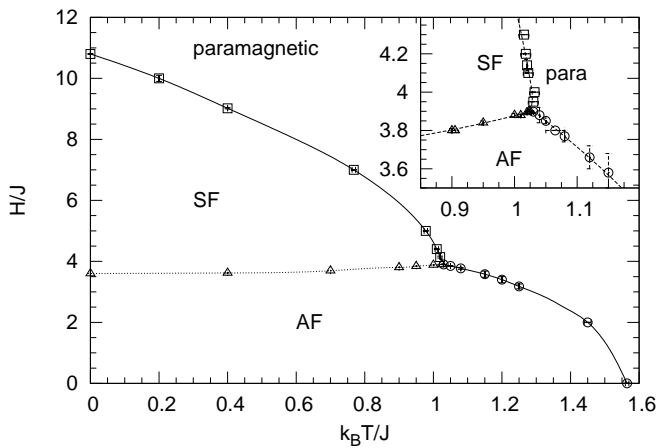}
 \caption{Phase diagram of the XXZ
  antiferromagnet on the cubic lattice, with $\Delta= 0.8$. Inset:
  Vicinity of the triple point.}
 \label{eps5}
\end{figure}

The nature of the triple point, at which the AF, SF, and P phases
meet, has been subject to recent renormalization group (RNG)
analyses, yielding either some kind of
critical endpoint \cite{vica} or, possibly depending on the 
degree of anisotropy $\Delta$, a bicritical point \cite{folk}. The
latter variant had been suggested in early RNG
studies \cite{mef} as well.   

From our MC simulations \cite{ban} we can locate the triple
point accurately at $k_BT/J= 1.025 \pm 0.015$
and $H/J= 3.90 \pm 0.03$, improving the previous
estimate \cite{bind}. We observe no evidence for a first--order
transition at the SF--P or AF--P boundary lines in the vicinity
of the triple point at $\Delta= 0.8$ \cite{ban}, being
compatibel with the existence of a bicritical point. In fact, we
find critical exponents, for instance for
staggered longitudinal (i.e. along the easy axis) and transversal
susceptibilities, at the AF--P boundary line to belong
to the Ising universality class, while those at the SF--P boundary
line are clearly consistent with the XY universality class. Perhaps, Monte
Carlo studies at different values of the exchange
anisotropy $\Delta$ may provide further insights to clarify the
predictions of the RNG treatments.

\section{Adding single--ion anisotropies}

There is no thermally stable biconical phase in
the classical XXZ antiferromagnet. BC configurations occur in the
ground state only at the special field, $H_c$, separating AF
and SF structures. BC fluctuations seem to lead
to a narrow disordered phase between the AF and SF
phases in two dimensions. In three (and, presumably,
higher) dimensions they do not destroy the phase
transition of first order between the AF and SF phases, as
predicted by mean--field theory. 

However, BC structures may be stabilized over a wide range of
fields, when introducing, for example, in the XXZ antiferromagnet
additional quadratic or cubic single--ion anisotropies
favoring non--uniaxial spin orientations. In the following we 
shall consider first phase diagrams, in the
($T$, $H$) plane, for classical Heisenberg
antiferromagnets with quadratic anisotropies
on square lattices 
and with cubic anisotropies on cubic lattices. Finally, we
shall present results on the ground state phase
diagram of a quantum, $S=1$, Heisenberg chain with uniaxial exchange
anisotropy plus a quadratic single--ion
anisotropy in a field.

We use straightforward ground state considerations and Monte Carlo
techniques for the classical models. The quantum spin chain at
zero temperature is analysed by applying density 
matrix renormalization group (DMRG) methods. Whenever feasible, the
finite--size behavior is taken into account.

In our MC simulations we study systems with up to $240^2$ spins 
in two dimensions, and with up to $32^2$ spins in three
dimensions. Details on the ground state DMRG calculations
for the $S=1$ quantum Heisenberg chain will be given below.

\subsection{Classical antiferromagnets in two and three dimensions}

In the case of the {\it square lattice}, we add to the XXZ model, eq. (1),
a single--ion term of the quadratic form

\begin{equation}
{\cal H}_D = D \sum\limits_{i} (S_i^z)^2
\end{equation}

\noindent
where the single--ion anisotropy may, depending on the sign of $D$,
enhance the uniaxial exchange anisotropy $\Delta$ ($0 < \Delta <
1$), when $D < 0$, or it may introduce a competing planar
anisotropy, $D > 0$. The full Hamiltonian is then
$\cal H= {\cal H}_{\mathrm{XXZ}} +{\cal H}_D$.

\begin{figure}[h]
  \includegraphics[width=.5\textwidth]{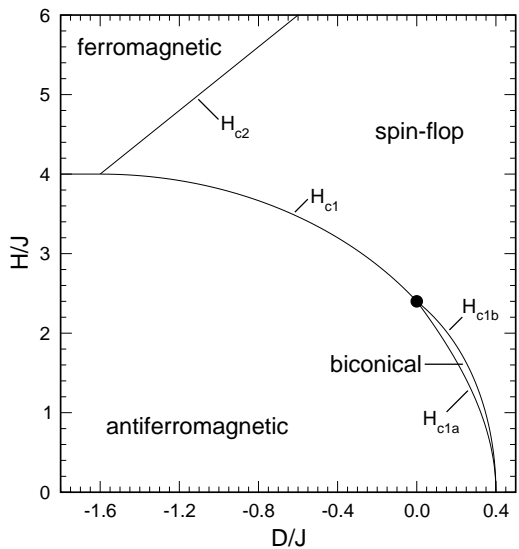}
 \caption{Ground state phase diagram of the classical Heisenberg model
  with exchange, $\Delta$, and quadratice single--ion, $D$, anisotropy on
  a square lattice, for $\Delta = 0.8$. The filled
  circle denotes the highly degenerate point at $D=0$ and $H=2.4 J$.}
\label{eps6}
\end{figure}

From a straighforward analysis of the ground states one may obtain
the phase diagram at zero temperature \cite{tsu,holt3,holt4}. An example
is shown in 
Fig. 6. There the exchange anisotropy has been fixed to be
$\Delta$= 0.8. The resulting ground state structures are of
AF, SF or BC type. Of course, at
sufficiently large fields, all spins will eventually point
in the direction of the field, i.e. one encounters the ferromagnetic
spin configuration, giving rise to the P phase at non--zero
temperatures. As for the XXZ antiferromagnet, the
xy--components of the spins order antiferromagnetically in the SF and
BC configurations, as depicted in Fig. 1, having rotational 
symmetry.

\begin{figure}[h]
  \includegraphics[width=.50\textwidth]{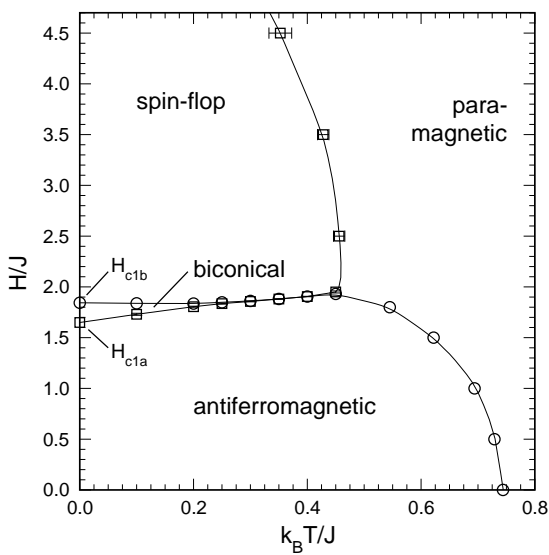}
\caption{Phase diagram of the XXZ antiferromagnet on a square 
         lattice with a competing single--ion
         anisotropy, $\Delta= 0.8$ and $D/J= 0.2$.}
\label{eps7}
\end{figure}

The most interesting feature of the ground state phase
diagram is the region of stable BC structures, evolving, at $D > 0$,
from the highly degenerate point of the XXZ model. In that region, the
degeneracy is lifted, and the tilt angles $\Theta_A$ and
$\Theta_B$, being still uniquely interrelated, change continuously 
with the field $H$ \cite{holt3,holt4}. Above a critical positive
value of $D$, $D/J > 2 -2\Delta$, at vanishing field, the spins
are oriented perpendicular to the easy axis. Applying then a field, a
SF structure will be stable, as shown in Fig. 6.

At $D>0$, the BC ground states give rise to a stable biconical
phase in between the AF and SF phases, as exemplified in
Fig. 7, setting $D/J= 0.2$. In the BC phase, the order
parameters of the AF and SF phases, the staggerd longitudinal
and transversal magnetizations, do not
vanish \cite{aha,tsu,liu,holt4}. The BC--AF transition is expected
to belong to the XY (or Kosterlitz--Thouless) universality
class, while at the BC--SF transition the order parameter of the
AF phase vanishes in an Ising--like manner.  

In the BC phase, the dominant tilt angles $\Theta_A$ and
$\Theta_B$ change continuously with the field, when fixing
the temperature. That behavior may be inferred from the 
ground state properties. It has been confirmed in simulations
by monitoring the probability $p(\Theta)$ of encountering a
spin with the tilt angle $\Theta$ at an arbitrary
lattice site, showing two pronounced peaks corresponding
to the values of $\Theta_A$ and $\Theta_B$ at given 
temperature and field in the BC phase \cite{holt4}.

It seems an open problem whether the AF, SF, BC, and P phases
meet at one point, presumably a tetracritical point \cite{bruce}. At
present, we can locate it, e.g. for the case depicted in
Fig. 7, only with a rather large uncertainty \cite{holt3,holt4}.

At $D<0$, BC structures are squeezed out by the single--ion
term in the ground state. At low temperatures, there seems to be
a direct transition of first order between the AF and SF
phases \cite{holt3,holt4}. 

In the case of the {\it cubic lattice}, we add to the XXZ model,
eq. (1), a cubic single--ion term \cite{ban}

\begin{equation}
  {\cal H}_{\mathrm{CA}} =  F \sum\limits_{i} \left[(S_i^x)^4 + (S_i^y)^4+
(S_i^z)^4 \right]
\end{equation}

\noindent
where $F$ denotes the strength of the cubic anisotropy. The sign of
$F$ determines whether the spins tend to align along the cubic
axes, for $F < 0$, case 1, or, for $F > 0$, case 2, in the
diagonal directions of the lattice. Because
of these tendencies, the BC, (i.e. BC1 or BC2), structures, as well as the
SF structures, show no full rotational invariance
in the $xy$--plane, in contrast
to the XXZ antiferromagnet, with or without quadratic single--ion
term. Now, as sketched in Fig. 8, the
discretized spin projections in the $xy$--plane
favor four directions \cite{ban}.

\begin{figure}[h]
  \includegraphics[width=.55\textwidth]{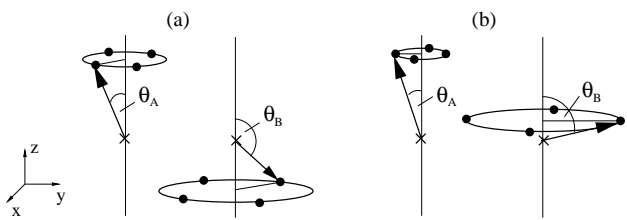}
\caption{Sketch of discretized biconical structures for the
    Heisenberg antiferromagnet with cubic anisotropy, in the case
    of (a) $F >0$, BC2, and (b)
    $F<0$, BC1. }
\label{eps8}
\end{figure}

The resulting ground state phase diagram of the
full Hamiltonian, ${\cal H}= {\cal H}_{\mathrm{XXZ}} + {\cal H}_{\mathrm{CA}}$, with
fixed exchange anisotropy, $\Delta =0.8$, and varying cubic term, $F$,
may be determined numerically
without difficulty \cite{ban,dinh}, as depicted in Fig. 9.

\begin{figure}[h]
  \includegraphics[width=.50\textwidth]{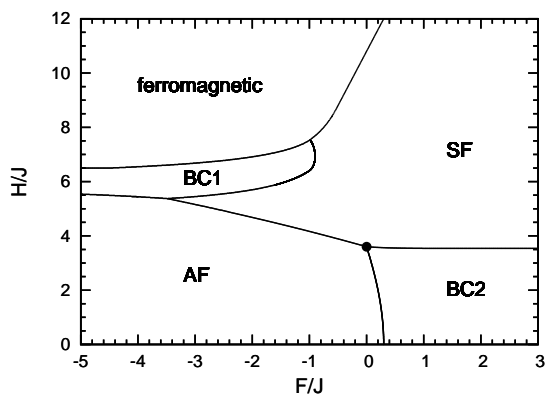}
 \caption{\label{label}Ground states of the
    three--dimensional classical XXZ antiferromagnet
    with exchange anisotropy, $\Delta= 0.8$, and
    varying the cubic term $F$.}
\label{eps9}
\end{figure}

For $F <0$, the transitions to the BC1 structures are typically
of first order, with a jump in the tilt angles, $\Theta$, with respect
to the $z$-axis, characterizing the BC
configurations. However, in the reentrance region between
the SF and BC1 structures at $F/J$ close to -1, the change in
the tilt angles seems to be smooth \cite{ban}.

Obviously, at non-zero temperatures, several interesting scenarios
leading, possibly, to multicritical behavior, where
AF, SF, BC, and P phases meet, may exist. So far, we focussed
attention on two cases \cite{ban}: 

(i) Positive cubic anisotropy $F > 0$, at
constant field $H/J = 1.8$ \cite{ban}, see Figs. 9 and 10. At small
values of $F$, there is an
AF ordering at low temperatures. Above a critical
value, $F_c =0.218...J$, the low--temperature phase is of BC2 type,
followed by the AF and P phases, when increasing the
temperature. The transition between the AF and P phases is found to
belong to the Ising universality class, while the transition between
the BC2 and AF phases seems to belong to the XY universality class,
with the cubic term being then an irrelevant
perturbation \cite{ban,vipe}. 

\begin{figure}[h]
  \includegraphics[width=.55\textwidth]{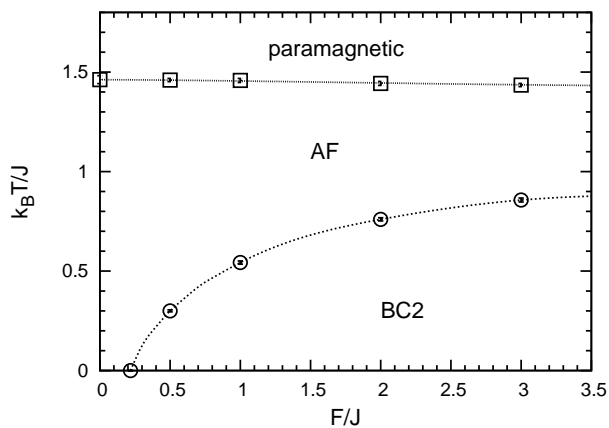}
\caption{Phase diagram of the XXZ antiferromagnet, $\Delta= 0.8$, on
        a cubic lattice with a positive cubic anisotropy $F$ at
        fixed field, $H/J$ =1.8.}
\label{eps10}
\end{figure}

(ii) Negative $F$, fixing the cubic
term, $F/J = -2$, and varying the field, see Fig. 9. In accordance with
the ground state analysis, we observe, at sufficiently low
temperatures, AF, SF, BC1, and P phases when increasing
the field. Interestingly, the BC1 phase
seems to become unstable when raising the temperature, with the other
phases being still present \cite{ban}. This may suggest that the
three boundary lines between the BC1--P, SF--BC1, and
SF--P phases meet at a multicritical point. Of course, further
clarification and a search for other, possibly multicritical
scenarios at different strengths of the cubic term, $F$, are
encouraged.

\subsection{Ground state phase diagrams of S=1 anisotropic Heisenberg chains}

In the following, we shall consider the spin--1 anisotropic
antiferromagnetic Heisenberg chain with a quadratic
single--ion term in a field described by the Hamiltonian

\begin{equation}
{\cal H} = \sum\limits_{i}
   (J (S_i^x S_{i+1}^x + S_i^y S_{i+1}^y
    + \Delta S_i^z S_{i+1}^z) + D (S_i^z)^2  - B S_i^z)
\end{equation}

\noindent
where $i$ denotes the lattice sites. The antiferromagnetic
exchange coupling $J$ is again positive. For $\Delta >1$, there is an
uniaxial exchange anisotropy, along the direction of the
field, $B >0$, the $z$--axis. Note the slight change 
in defining the exchange anisotropy $\Delta$ compared
to that for the XXZ model as given in eq. (1). We here follow
the standard notation of previous studies on closely related
quantum spin chains \cite{sak,sen,peters}. For the same
reason, the field is now denoted by $B$. Depending on the
sign of $D$, the single-ion term leads to a competing
planar anisotropy, $D > 0$, or to an enhancing uniaxial anisotropy, $D<0$, as
stated already above. We shall analyse ground
state properties using DMRG techniques \cite{white,scholl} for chains with
open boundary conditions and up to $L= 128$ sites. In addition, we
determine the ground states of corresponding infinite
chains with classical spin vectors of length one \cite{tsu,holt4}.

We shall compare our findings on the quantum spin chain to
ground state properties of the corresponding classical chains,
eq. (5). To characterise the quantum
spin structures, the analogue of the classical spin--flop
configuration will be called ``spin--liquid' (SL) structure, and the 
analogue of the classical biconical configuration will be
called supersolid (SS) structure, following the actual standard
terminology.
 
We first briefly deal with the case $D/J= \Delta/2$
\cite{sen,peters}. The ground state phase diagram in the
$(\Delta,B/J)$ plane has been found \cite{sen,peters} to comprise
AF, SL, SS, ferromagnetic (F), and (10), with a magnetization plateau
at half saturation, phases. At small values of $\Delta$ and
small fields, the Haldane phase is observed \cite{sen,haldane}. In
comparison, the
corresponding classical spin chain shows a much broader
biconical (BC) phase, being effectively replaced not only by the 
supersolid phase but also, largely, by the SL and (10)
phases. The classical (10) phase becomes stable only in the limit of an
Ising antiferromagnetic chain with a single--ion
term, the Blume--Capel model \cite{peters}.

\begin{figure}[h]
  \includegraphics[width=.55\textwidth]{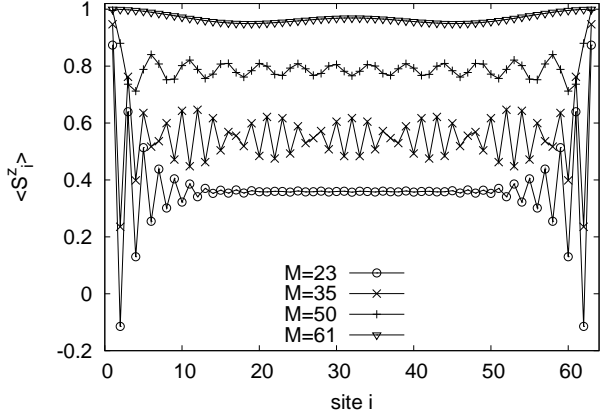}
\caption{Magnetization profiles
in the spin--liquid phase for $\Delta =2D/J= 3.5$ at various total
magnetizations for the $S=1$ chain, eq. (5), with $L= 63$ sites.}
\label{eps11}
\end{figure}

Interesting information is given by the
magnetization profiles, $m_i= \langle S^z_i \rangle$, with
brackets, $\langle ...\rangle$,
denoting quantum mechanical expectation values. In particular, we
observe distinct profiles in the SS phase in between the AF and
(10) and in between the AF and SL  phases, respectively. For odd
$L$, in the SS phase on approach to the (10) phase, the local magnetizations
$m_i$ at odd sites stay close to one, while at even sites they tend
roughly to zero. In
contrast, in the SS phase on approach to the SL phase, the
magnetizations on odd and even sites
tend to take on the same values \cite{peters}.

In the SL phase we
find also (studying situations with $\Delta$ exceeding $\approx$ 2.5)
two distinct types of profiles, for finite chains, when
varying $M/L$, where $M$ is
the total magnetization \cite{peters}: For $M < L/2$, the
profiles exhibit a broad plateau in
the center of the chain, as expected for a classical spin--flop (SF) 
structure, while pronounced modulations in $m_i$
occur at $M >L/2$, see Fig. 11. This may signal a change from commensurate (C)
to incommensurate (IC) structures \cite{peters}, with
an interesting quantum phase transition. The suggestion is
confirmed and quantified by analysing the Fourier transform
of the profiles, especially at $\Delta= 3.5$. The
modulation in the IC region of the
SL phase is described nicely by the wavenumber (setting the lattice
spacing equal to one) $q= 2 \pi (1-m)$, $m= M/L$. Such
an IC modulation, with algebraic decay, is expected to
hold in the entire SL phase of the spin--1/2 anisotropic Heisenberg
chain in a field \cite{kimu}. In the classical variant, we find no
IC structures in the SF phase, for finite and
infinite chains.

\begin{figure}[h]
  \includegraphics[width=.55\textwidth]{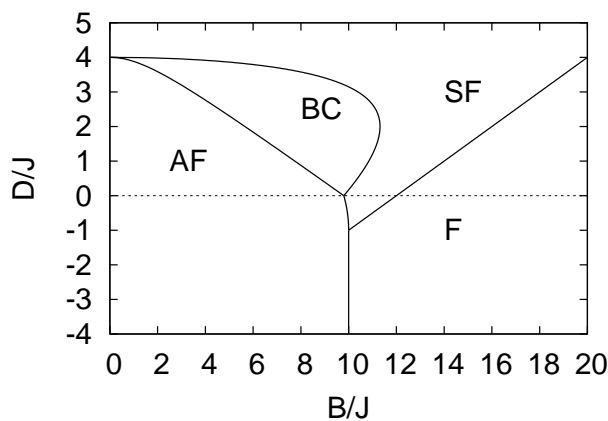}
\caption{\label{label} Ground state phase diagram of the classical
  infinite anisotropic Heisenberg chain, eq. (5), with $\Delta= 5.0$.}
  \label{eps12}
\end{figure}

Let us now turn to the case of fixed exchange
anisotropy, $\Delta= 5.0$, varying the
single--ion term, $D$. The ground state phase diagrams for
the classical and the spin--1 chains are depicted in Figs. 12 and
13, using DMRG calculations for chains with up to 63 sites
for the quantum case, and straightforward ground state
considerations \cite{holt4} (checked by
Monte Carlo data) for the infinite classical chain. We considered
positive and negative single--ion
anisotropies, $D$, finding, especially, several intrigiung quantum
phase transitions. 

\begin{figure}[h]
 \includegraphics[width=.55\textwidth]{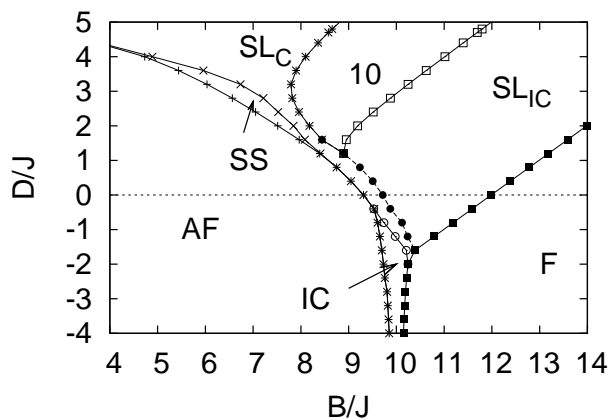}
\caption{\label{label}Ground state phase diagram of the spin--1
  anisotropic Heisenberg chain, eq. (5), with $\Delta= 5.0$.}
\label{eps13}
\end{figure}

For $D > 0$, BC and SS phase are stable at
zero temperature, see Figs. 12 and 13. As in the
case of $D/J= \Delta/2$, the broad BC phase
of the classical chain is effectively replaced, in the quantum chain,
by the corresponding, rather narrow SS phase as well as
SL and (10) phases.

In contrast to the case $D/J= \Delta/2$, the supersolid phase
is always bordered by the AF and SL phases. Accordingly, we
observe only one type of magnetization profile. Illustrative
examples are depicted in Fig. 14, at $D/J= 3.0$ and various
fields. In the SS phase, at given single--ion anisotropy, $D$, and
field, $B$, the magnetization takes on different values
at odd and even sites in the center of the
chain. The local magnetization $m_i$ tends to acquire the same
value at odd and even sites on approach to the SL phase in the quantum
chain. Actually, as described above, the classical BC phase
is characterized by two tilt angles $\Theta_A$ and $\Theta_B$ for
the two sublattices formed by neighboring sites, with the tilt angles
approaching each other when getting closer to the SF
phase \cite{holt2,holt4}. Obviously, this behavior is
completely analogous to the one depicted in Fig. 14.

\begin{figure}[h]
  \includegraphics[width=.55\textwidth]{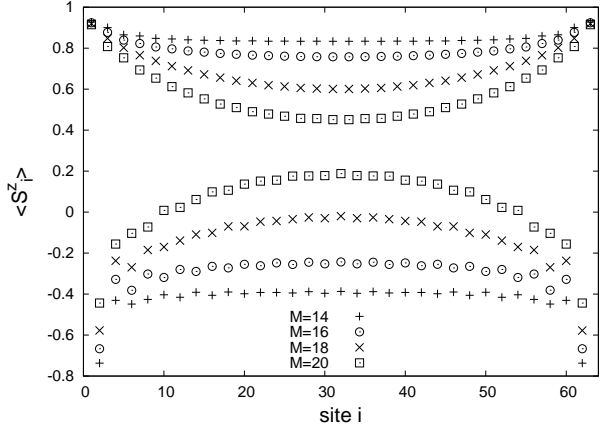}
\caption{Magnetization profiles
in the supersolid phase at $\Delta= 5.0$, $D/J= 3.0$, and fields $B/J$
(corresponding to total magnetizations $M$) between 6.5 and 6.8. Spin
chains with $L= 63$ sites
are considered. The lower (upper) symbols denote the even (odd) sites.}
\label{eps14}
\end{figure}

Increasing $D$, the 'large--D'
phase \cite{schulz,mik,chen,oitmaa} may eventually be
stable, leading to other intrigiung quantum
phase transitions. It corresponds to the
planar phase in the classical model with vanishing field, with the spin vectors
pointing perpendicular to the $z$--axis, being the ground state
for $D/J \ge 4$. The new phase may be
expected to give rise to a SL phase at
non--zero fields. A discussion
of this interesting aspect is, however, beyond the scope of the present
contribution.

For $D< 0$, biconical and supersolid phase are no
longer stable. In fact, only AF, SL (or SF),
and F structures are encountered, in accordance with the
discussion in the preceeding subsection. The SL phase can be either
commensurate, with a wide plateau in the magnetization profile
away from the boundaries, or incommensurate, with modulations
superimposed on the average magnetization. Obviously, there
are two IC phases, see Fig. 13. The one, denoted by 'IC' in
that figure, occurring essentially in between
the AF and F phases, has been found before, having
exponentially decaying transversal spin--spin
correlations \cite{sak}, in contrast to the usual spin--liquids
with algebraic decay. It has no analog in the classical model, see
Fig. 12. The related transition between the IC and
SL$_C$  phases has been obtained before, being either of first
order or, at large average
magnetization, continuous \cite{sak}. Our results agree with
that description. We find another C--IC transition
line between the SL$_C$ and SL$_{IC}$ phases at somewhat larger fields
for given $D/J$, see Fig. 12 (full circles). Note that this
transition seems to take place at $M/L$ significantly larger
than 1/2 for $D <0$. Increasing $D$, $D> 0$, in the vicinity of
the (10) phase, the line goes over to the
above discussed scenario with commensurate, $M/L < 1/2$,  and
incommensurate, $M/L > 1/2$, structures. By further
increasing the
planar single--ion anisotropy, $D/J$ being larger than
roughly 3.6, we observe in the 'SL$_C$' phase
close to the (10) phase modulated structures for rather
short chains, $L \le 31$. The possible, additional C--IC border is not
displayed in Fig. 13. Indeed, a more detailed analysis of
that region, taking into
account finite--size effects, is desirable.

\section{Summary}

In this contribution, we have presented results of recent and
current studies on a 
variety of anisotropic Heisenberg antiferromagnets in a 
field, elucidating the role of exchange and crystal field
single--ion anisotropies, of lattice dimension, and
of quantum effects.

The prototypical classical XXZ Heienberg antiferromagnet with
uniaxial exchange anisotropy in a field along the easy axis is shown
to exhibit highly degenerate biconical ground states at the field
separating AF and SF configurations. The BC structures give rise
to a narrow disordered phase intervening between the AF and SF
phases at low temperatures for square lattices. In contrast, for
cubic lattices the BC fluctuations do not destroy the direct
transition of first order between the two ordered phases. Quantum
effects seem to suppress the BC structures in the S=1/2 XXZ model
on a square lattice, leading, presumably, to a transition of first
ortder between the AF and SF phases.

Thermally stable BC phases may occur when adding to
the exchange anisotropy of the XXZ antiferromagnet, for instance, a
quadratic or cubic single--ion anisotropy favoring a non--uniaxial
ordering of the spins. Resulting phase diagrams for classical two-- and
three--dimensional magnets exhibit intriguing multicritical 
points.

Studying ground state properties of $S=1$ anisotropic
Heisenberg chains with an additional quadratic single--ion
anisotropy, we observe, for finite chains, two distinct
types of spin--liquid as well as two distinct
types of supersolid structures. The region of stability of
the supersolid structures is substantially reduced, compared
to that of the corresponding biconical structures in the classical
case. Of course, there are no thermal phase transitions in
one--dimensional magnets with short--range
interactions. However, there are interesting quantum phase transitions
at zero temperature.

\section*{Acknowledgements}

We should like to thank, especially,  A. Aharony, C. D. Batista,
 A. N. Bogdanov, T.--C. Dinh, R. Folk,  A. Kolezhuk, N. Laflorencie,
 D. P. Landau, R. Leidl, P. Sengupta, and J. Sirker for
useful discussions and information. A part of the research has
been funded by the excellence initiative of
the German federal and state governments.

\end{document}